\RequirePackage[2020-02-02]{latexrelease}
\documentclass[times, twoside]{zHenriquesLab-StyleBioRxiv}
\usepackage{color,soul}

\usepackage{placeins}
\newcommand{\code}[1]{\texttt{#1}}
\usepackage{lineno}
\usepackage[utf8]{inputenc} 
\usepackage[T1]{fontenc}    
\usepackage{booktabs}       
\usepackage{amsfonts}       
\usepackage{nicefrac}       
\usepackage{microtype}      
\usepackage{xcolor}         
\usepackage{ulem}
\usepackage{color,soul}
\usepackage{placeins}
\usepackage{graphicx}
\usepackage{footmisc}
\usepackage{multirow}
\usepackage{amssymb}

\def\sensorium{\texttt{SENSORIUM}}
\def\sensoriump{\texttt{SENSORIUM+}}

\leadauthor{Willeke*, Fahey*} 

\begin{document}


\title{The Sensorium competition on predicting large-scale mouse primary visual cortex activity}
\shorttitle{\sensorium~ 2022}
 
\author[1-3,*, \Letter]{Konstantin F. Willeke}
\author[4,5,*, \Letter]{Paul G. Fahey}
\author[1-3]{Mohammad Bashiri}
\author[3]{Laura Pede}
\author[1,3,6]{Max F. Burg}
\author[3]{Christoph Blessing}
\author[1,3,6]{Santiago A. Cadena}
\author[4,5]{Zhiwei Ding}
\author[1-3]{Konstantin-Klemens Lurz}
\author[4,5]{Kayla Ponder}
\author[4,5]{Taliah Muhammad}
\author[4,5]{Saumil S. Patel}
\author[3,7]{Alexander S. Ecker}
\author[4,5,8]{Andreas S. Tolias}
\author[2-5]{Fabian H. Sinz}

\affil[1]{International Max Planck Research School for Intelligent Systems, University of Tübingen, Germany}
\affil[2]{Institute for Bioinformatics and Medical Informatics, University of Tübingen, Germany}
\affil[3]{Institute of Computer Science and Campus Institute Data Science, University of Göttingen, Germany}
\affil[4]{Department of Neuroscience, Baylor College of Medicine, Houston, TX, USA}
\affil[5]{Center for Neuroscience and Artificial Intelligence, Baylor College of Medicine, Houston, TX, USA}
\affil[6]{Institute for Theoretical Physics, University of Tübingen, Germany}
\affil[7]{Max Planck Institute for Dynamics and Self-Organization, Göttingen, Germany}
\affil[8]{Electrical and Computer Engineering, Rice University, Houston, USA}
\affil[*,]{Equal contributions}

\maketitle

\begin{abstract}

The neural underpinning of the biological visual system is challenging to study experimentally, in particular as the neuronal activity becomes increasingly nonlinear with respect to visual input. 
Artificial neural networks (ANNs) can serve a variety of goals for improving our understanding of this complex system, not only serving as predictive digital twins  of sensory cortex for novel hypothesis generation \textit{in silico}, but also incorporating bio-inspired architectural motifs to progressively bridge the gap between biological and machine vision.  
The mouse has recently emerged as a popular model system to study visual information processing, but no standardized large-scale benchmark to identify state-of-the-art models of the mouse visual system has been established. 
To fill this gap, we propose the \sensorium~ benchmark competition.
We collected a large-scale dataset from mouse primary visual cortex containing the responses of more than 28,000 neurons across seven mice stimulated with thousands of natural images, together with simultaneous behavioral measurements that include running speed, pupil dilation, and eye movements.
The benchmark challenge will rank models based on predictive performance for neuronal responses on a held-out test set, and includes two tracks for model input limited to either stimulus only (\sensorium) or stimulus plus behavior (\sensoriump).
We provide a starting kit to lower the barrier for entry, including tutorials, pre-trained baseline models, and APIs with one line commands for data loading and submission.
We would like to see this as a starting point for regular challenges and data releases, and as a standard tool for measuring progress in large-scale neural system identification models of the mouse visual system and beyond.

\end{abstract}

\begin{corrauthor}
konstantin-friedrich.willeke\at uni-tuebingen.de; paul.fahey\at bcm.edu; sinz@cs.uni-goettingen.de
\end{corrauthor}

\subsection*{Keywords}
mouse visual cortex, system identification, neural prediction, natural images

\begin{figure*}[ht!]
    \begin{minipage}[t]{0.95\linewidth}
        \vspace{0 pt}
        \includegraphics[trim=0 0 0 0, clip,width=\linewidth]{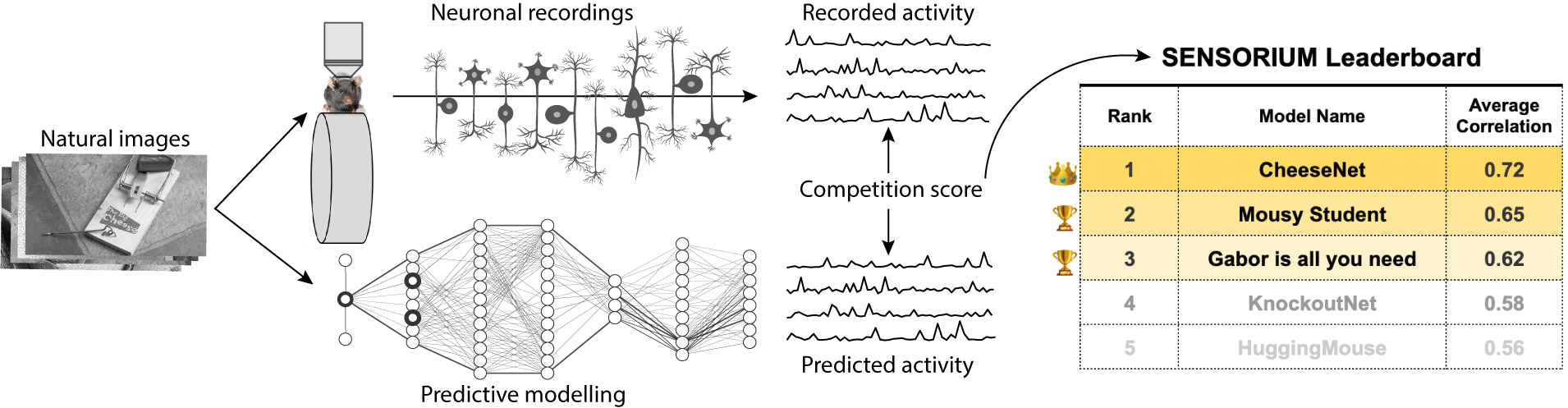}
    \end{minipage}\hfill
    \begin{minipage}[t]{.95\linewidth}
        \vspace{20pt}
        \caption{\textbf{A schematic illustration of the \sensorium~ competition.}
             We will provide large-scale datasets of neuronal activity in the primary visual cortex of mice. Participants of the competition will train models on pairs of natural image stimuli and recorded neuronal activity, in search for the best neural predictive model. \label{fig:Fig_1}
        }
    \end{minipage}
\end{figure*}

\section*{Introduction}

Understanding how the visual system processes visual information is a long standing goal in neuroscience. 
Neural system identification approaches this problem in a quantitative, testable, and reproducible way by building accurate predictive models of neural population activity in response to arbitrary input.
If successful, these models can serve as phenomenological digital twins for the visual cortex, allowing computational neuroscientists to derive new hypotheses about biological vision \textit{in silico}, and enabling systems neuroscientists to test them \textit{in vivo} \citep{Walker2019-oq, Ponce2019-yn, Bashivan2019, Franke2021}. 
In addition, highly predictive models are also relevant to machine learning researchers who use them to bridge the gap between biological and machine vision \citep{Li2019-il,Safarani2021-yy,Li2022-ot,Sinz2019}. 

The work on predictive models of neural responses to visual inputs has a long history that includes simple linear-nonlinear (LN) models \citep{Jones1987-sn,Heeger1992-ig, Heeger1992-xx}, energy models \citep{Adelson1985-re}, more general subunit/LN-LN models \citep{Rust2005,Touryan2005,Schwartz2006-ji, Vintch2015}, and multi-layer neural network models \citep{Zipser1988,Lehky1992,Lau2002,Prenger2004}. 
The deep learning revolution set new standards in prediction performance by leveraging task-optimized deep convolutional neural networks (CNNs) \citep{Yamins2014,Cadieu2014,Cadena2019} and CNN-based architectures incorporating a shared encoding learned end-to-end for thousands of neurons \citep{Antolik2016,Batty2016,McIntosh2016,Klindt2017,Kindel2017,Cadena2019,Burg2021, Lurz2020-ua, Bashiri2021-or,Zhang2018-cs,Cowley2020neurips,Ecker2018, Sinz2018-sk, Walker2019-oq, Franke2021}.

The core idea of a neural system identification approach to improve our understanding of an underlying sensory area is that models that explain more of the stimulus-driven variability may capture nonlinearities that previous low-parametric models have missed \citep{carandiniWeKnowWhat2005}. Subsequent analysis of high performing models, paired with ongoing \textit{in vivo} verification, can eventually yield more complete principles of brain computation.  This motivates continually improving our models to explain as much as possible of the stimulus-driven variability and analyze these models to decipher principles of brain computations. 

Standardized large-scale benchmarks are one approach to stimulate constructive competition between models compared on equal ground, leading to numerous seemingly small and incremental improvements that accumulate to substantial progress. In machine learning and computer vision, benchmarks have been an important driver of innovation in the last ten years. For instance, benchmarks such as the ImageNetChallenge \citep{Russakovsky2015-xi} have helped jump start the revolution in artificial intelligence through deep learning. 



Similarly, neuroscience can benefit from more large-scale benchmarks to drive innovation and identify state-of-the-art models. This is especially true in the mouse visual cortex, which has recently emerged as a popular model system to study visual information processing, due to the wide range of available genetic and light imaging techniques for interrogating large-scale neural activity. 

Existing neuroscience benchmarks vary substantially in the type of data, model organism, or goals of the contest \citep{Schrimpf2018, Algonauts2021, deVries2019, latentsbench}.  For example, the \textbf{Brain-Score} benchmark \citep{Schrimpf2018} ranks \textit{task}-pretrained models that best match areas across primate visual ventral stream and other behavioral data, but rather than providing neuronal training data, participants are expected to design objectives, learning procedures, network architectures, and input data that result in representations that are predictive of the withheld neural data.  The \textbf{Algonauts} challenge \citep{Algonauts2021} competition ranks neural predictive models of human brain activity recorded with fMRI in visual cortex in response to natural images and videos.  Additionally, large data releases such as the extensively annotated dataset in mouse visual cortex from \textbf{Allen Institute for Brain Science} \citep{deVries2019} are often not designed for a machine learning competition (consisting of only 118 natural images in addition to parametric stimuli and natural movies), and lacking benchmark infrastructure for measuring predictive performance against a withheld test set.

To fill this gap, we created the \sensorium~ benchmark competition to facilitate the search for the best predictive model for mouse visual cortex. 
We collected a large-scale dataset from mouse primary visual cortex containing the responses of more than 28,000 neurons across seven mice stimulated with thousands of natural images, together with simultaneous behavioral measurements that include running speed, pupil dilation, and eye movements.
Benchmark metrics will rank models based on predictive performance for neuronal responses on a held-out test set, and includes two tracks for model input limited to either stimulus only (\sensorium) or stimulus plus behavior (\sensoriump).

We also provide a starting kit to lower the barrier for entry, including tutorials, pre-trained baseline models based on published architectures \citep{Klindt2017,Lurz2020-ua}, and APIs with one line commands for data loading and submission.
Our goal is to continue with regular challenges and data releases, as a standard tool for measuring progress in large-scale neural system identification models of the mouse visual hierarchy and beyond.

\begin{figure*}[ht!]
    \begin{minipage}[t]{1\linewidth}
        \vspace{0 pt}
        \centering
        \includegraphics[trim=0 0 0 0, clip,width=\linewidth]{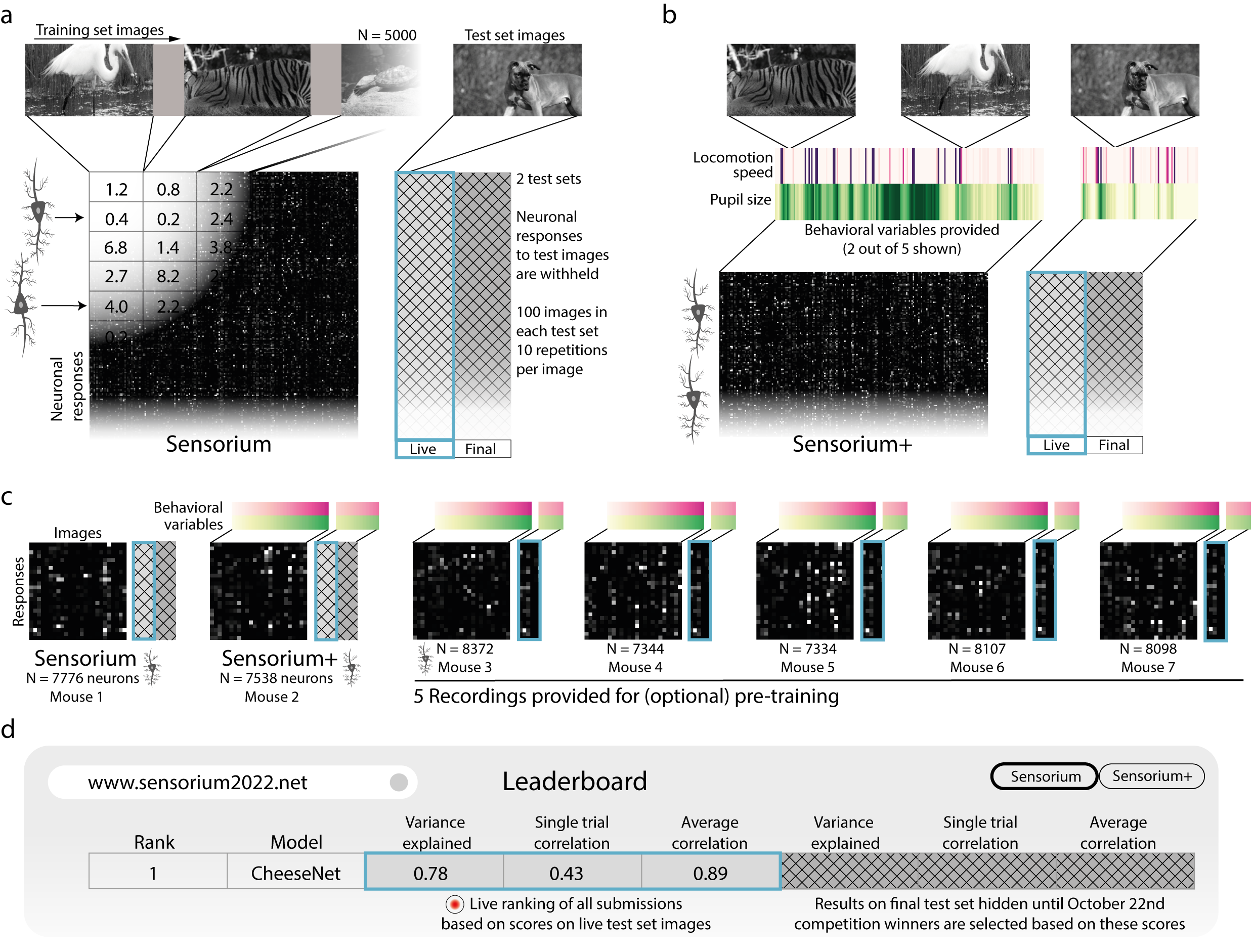}
    \end{minipage}\hfill
    \begin{minipage}[t]{0.95\linewidth}
        \vspace{10 pt}
        \caption{\textbf{Overview of the data and the competition structure.}
             \textbf{a}, Detailed view of the data format of a single recording used for the \sensorium~ track. The data consists of training images (N$\approx$5000) and the associated scalar neuronal activity of each neuron. Furthermore, there are two test sets, the \textit{live} and \textit{final} sets, consisting of 100 images, shown 10 times each to the animal. The neuronal responses to both test sets are withheld. \textbf{b}, \sensoriump~ track. Same as in (a) but for this track, the behavioral variables are also provided. \textbf{c}, overview of the seven recordings in our dataset. The \textit{pre-training} recordings are not part of the competition evaluation, but they can be used to improve model performance. The \textit{live test} set images are also contained in the pre-training recordings (see blue frame in all panels), along with the neuronal responses. Thus, we refer to this set as the \textit{public test} set. In summary: the \textit{live} and \textit{public test} set have the same images, but different neurons. \textbf{d}, the \textit{live test} set scores are displayed on the live leaderboard. The \textit{final test} set scores will be validated and revealed after the submissions close.  
             \label{fig:Fig_2}
        }
    \end{minipage}
\end{figure*}

\section*{The \sensorium \  Competition} 
The goal of the \sensorium~ 2022 competition is to identify the best models for predicting a large number of sensory neural responses to arbitrary natural stimuli.  
The start of the competition is accompanied by the public release of a training dataset for refining model performance (for more details, see Data and Fig.~\ref{fig:Fig_2}), including two animals for which a competition test set has been withheld. 

For the held-out competition test set, the recorded neuronal responses will not be (and have never been) publicly released.  The test set images are divided into two exclusive groups: \textit{live} and \textit{final test}.
Performance metrics (see Metrics) computed on the \textit{live test} images will be used to maintain a public leaderboard on our website throughout the submission period, while the performance metrics on the \textit{final test} images will be used to identify the winning entries, and will only be revealed after the submission period has ended (Fig.~\ref{fig:Fig_2}d).  
By separating the \textit{live test} and \textit{final test} set performance metrics, we are able to provide feedback on performance on the \textit{live test} set to participants wishing to submit updated predictions over the course of the competition (up to the limit of one submission per day), while protecting the validity of the \textit{final test} set from overfitting over multiple submissions.

The competition has two tracks, \sensorium~ and \sensoriump~, predicting two datasets with the same stimuli, but from two different animals and with differing model inputs, as detailed in the following.

\textbf{\sensorium.}
In the first challenge, participants have to predict neuronal activity of 7,776 neurons in response to 200 unique natural images of our competition \textit{live test} and \textit{final test} image sets.  
The data provided for the test set includes the natural image stimuli but not the behavioral variables (Fig.~\ref{fig:Fig_2}a).  
Thus, the focus of this challenge is the stimulus-driven responses, treating other correlates of neural variability, such as behavioral state, as noise.
This track resembles most of the current efforts in the community \citep{Schrimpf2018} to identify stimulus-response functions without any additional information about the brain and behavioral state.

\textbf{\sensoriump}
In the second challenge, participants will predict neuronal activity of 7,538 neurons in response to 200 unique natural images of our competition \textit{live test} and \textit{final test} image sets. 
In this case, the data provided includes both the natural image stimuli and the accompanying behavioral variables (Fig.~\ref{fig:Fig_2}b, see Sec.~\ref{sec:Data} below). 
As a significant part of response variability correlates with the animal's behavior and internal brain state \citep{niell2010modulation, Reimer2014-ry, stringer2019spontaneous}, their inclusion in the modeling process can result in models that capture single trial neural responses more accurately \citep{Bashiri2021-or, Franke2021}.


\section*{Data}
\label{sec:Data}

The competition dataset was designed with the goal of comparing neural predictive models that capture neuronal responses $\mathbf{r} \in \mathbb{R}^{n}$ of ${n}$ neurons as a function $\mathbf{f}_\theta(\mathbf{x})$ of either only natural image stimuli $\mathbf{x} \in \mathbb{R}^{h\times w}$ (image \textbf{h}eight,\textbf{w}idth), or as a function $\mathbf{f}_\theta(\mathbf{x}, \mathbf{b})$ of both natural image stimuli and behavioral variables $\mathbf{b} \in \mathbb{R}^{k}$.  We provide $k=5$ variables: locomotion speed, pupil size, instantaneous change of pupil size (second order central difference), and horizontal and vertical eye position. 
See Fig.~\ref{fig:Fig_2} for an overview of the dataset.

\textbf{Natural images.} We sampled natural images from ImageNet \citep{Russakovsky2015-xi}. The images were cropped to fit a monitor with 16:9 aspect ratio, converted to gray scale, and presented to mice for 500~ms, preceded by a blank screen period between 300 and 500~ms (Fig.~\ref{fig:Fig_2}a,b)

\textbf{Neuronal responses.} 
We recorded the response of excitatory neurons in layer 2/3 of the right primary visual cortex in awake, head-fixed, behaving mice using calcium imaging. Neuronal activity was extracted and accumulated between 50 and 550 ms after each stimulus onset using a boxcar window. 

\textbf{Behavioral variables.} During imaging, mice were head-mounted over a cylindrical treadmill, and an eye camera captured changes in the pupil position and dilation. Behavioral variables were similarly extracted and accumulated between 50 and 550 ms after each stimulus onset using a boxcar window. 

\textbf{Dataset.}
Our complete corpus of data comprises seven recordings in seven animals (Fig.~\ref{fig:Fig_2}c), which in total contain the neuronal activity of more than 28,000 neurons to a total of 25,200 unique images, with 6,000--7,000 image presentations per recording.
We report a conservative estimate of the number of neurons present in the data, due to dense recording planes in our calcium imaging setup, which leads to single neurons producing multiple segmented units by appearing in multiple recording planes. Fig.~\ref{fig:Fig_2}c shows the uncorrected number with a total of 54,569 units.

Five of the seven recordings, which we refer to as \textit{pre-training recordings} (Fig.~\ref{fig:Fig_2}c, right), are provided solely for training and model generalization. The pre-training recordings are not included in the competition performance metrics. They contain 5,000 single presentations of natural images which are randomly intermixed with 10 repetitions of 100 natural images, and all corresponding neuronal responses. The 100 repeated images and responses serve as a \textit{public test} set for each recording. 

In the two remaining recordings, our \textit{competition recordings} (Fig.~\ref{fig:Fig_2}c, left), the mice were also presented with 5,000 single presentations of training stimuli as well as the \textit{public test} images. However, during the contest, we withhold the responses to the \textit{public test} images and use them for the live leaderboard during the contest. We thus refer to these images as \textit{live test} set. Furthermore, the competition recordings contain 10 repetitions of 100 additional natural \textit{test} images that were randomly intermixed during the experiment. These \textit{test} images are only present in the two competition recordings. The responses to these images will also be withheld and used to determine the winner of the competition after submissions are closed (Fig.~\ref{fig:Fig_2}a,b). We refer to these images as our \textit{final test} set. By providing both \textit{live} and \textit{final test} scoring, participants receive the benefit of iterative feedback while avoiding overfitting on the final scoring metrics.

In our first competition track (\sensorium, Fig.~\ref{fig:Fig_2}a), we withhold the behavioral variables, such that only the natural images can be used to predict the neuronal responses. For the other competition track (\sensoriump, Fig.~\ref{fig:Fig_2}b), as well as the five pre-training recordings recordings, we are releasing all the behavioral variables.
For the pre-training recordings, we are also releasing the order of the presented images as shown to the animal, but we withhold this information in the competition datasets, to discourage participants from learning neuronal response trends over time. Lastly, we are releasing the anatomical locations of the recorded neurons for all datasets. 

The complete corpus of data is available to download \footnote[1]{\url{http://sensorium2022.net/}}.

\setlength{\tabcolsep}{5pt}
\renewcommand*\arraystretch{1.4}

\begin{table*}[t]
\centering
\begin{tabular}{lllllllll}
\hline
                            &              & \multicolumn{7}{c}{Baseline performance on held out test data}    \\ 
\hline
                            &              & \multicolumn{3}{c}{Live test set}&  & \multicolumn{3}{c}{Final test set}\\ 
\cline{3-5}\cline{7-9}
Competition track           &              & \multicolumn{1}{c}{\begin{tabular}[c]{@{}c@{}}Single Trial\\ Correlation\end{tabular}} & \multicolumn{1}{c}{\begin{tabular}[c]{@{}c@{}}Correlation\\ to Average\end{tabular}} & \multicolumn{1}{c}{FEVE} &  & \multicolumn{1}{c}{\begin{tabular}[c]{@{}c@{}}Single Trial\\ Correlation\end{tabular}} & \multicolumn{1}{c}{\begin{tabular}[c]{@{}c@{}}Correlation\\ to Average\end{tabular}} & \multicolumn{1}{c}{FEVE}  \\ 
\hline
\multirow{2}{*}{\sensorium}  & CNN Baseline & \multicolumn{1}{c}{.274}& \multicolumn{1}{c}{.513}& \multicolumn{1}{c}{.433}&  & \multicolumn{1}{c}{.287}& \multicolumn{1}{c}{\textbf{.528}}& \multicolumn{1}{c}{.439}\\
                            & LN Baseline  & \multicolumn{1}{c}{.197}& \multicolumn{1}{c}{.363}& \multicolumn{1}{c}{.222}&  & \multicolumn{1}{c}{.207}& \multicolumn{1}{c}{\textbf{.377}}& \multicolumn{1}{c}{.232}\\
                            &              & & & &  & & & \\
\multirow{2}{*}{\sensoriump} & CNN Baseline & \multicolumn{1}{c}{.374} & \multicolumn{1}{c}{.571} & \multicolumn{1}{c}{-} &  & \multicolumn{1}{c}{\textbf{.384}} & \multicolumn{1}{c}{.578} & \multicolumn{1}{c}{-} \\
                            & LN Baseline  & \multicolumn{1}{c}{.257} & \multicolumn{1}{c}{.373} & \multicolumn{1}{c}{-} &  & \multicolumn{1}{c}{\textbf{.266}} & \multicolumn{1}{c}{.385} & \multicolumn{1}{c}{-} \\
\hline
\end{tabular}
\caption{
        \textbf{Performance of the baseline models on both tracks of the competition.}
        For the start of our competition, we are providing baseline scores for two models in each competition track. The main score which will determine the winner of the competition is written in boldface. Note that the fraction of explainable variance explained (FEVE) is calculated on a subset of neurons with an explainable variance larger than 15\% (See section Materials and Methods for details). Furthermore, FEVE can not be computed for the \sensoriump~ track (see Materials and Methods).
        \label{table:T1}
        }
\end{table*}

\section*{Performance Metrics}
\label{sec:Metrics}

Across the two benchmark tracks, three metrics of predictive accuracy will be automatically and independently computed for the 100 \textit{live test} set images and 100 \textit {final test} set images, for which the ground-truth neuronal responses are withheld.  
Due to their nature, not all three metrics will be computed for both tracks of the competition and only one of the three metrics on the {final test} set will be used for independently ranking the submissions to the two benchmark tracks. See Materials and Methods for further details and formulae.

\textbf{Correlation to Average}
We calculate the \textit{correlation to average} $\rho_{\textrm{av}}$ of 100 model predictions to the withheld, observed mean neural response across 10 repeated presentations of the same stimulus.
This metric will be computed for both the \sensorium~ and \sensoriump~ tracks to facilitate comparison.
Correlation to average on the \textit{final test} set will serve as the ultimate ranking score in the \sensorium~ track to determine competition winners. 

\textbf{Fraction of Explainable Variance Explained (FEVE)}
While the Correlation to average is a common metric, it is insensitive to affine transformations of either the neuronal response or predictions.
Thus, we also compute FEVE to measure how well the model accounts for stimulus-driven (i.e. explainable) variations in neuronal response \citep{Cadena2019}.
This metric computes the ratio between the variance explained by the model and the explainable variance in the neural responses.  
The explainable variance accounts for only the stimulus-driven variance and ignores the trial-to-trial variability in responses (see Materials and Methods for details).
This metric will be computed for \sensorium~ but not \sensoriump~, due to the lack of repeated trials with the exactly same behavior fluctuations necessary to estimate the noise ceiling with FEVE. For numerical stability, we compute the FEVE only for neurons with an explainable variance larger than 15\% (See section Materials and Methods for details).

\textbf{Single Trial Correlation}
Lastly, to measure how well models account for trial-to-trial variations we compute the \textit{single trial correlation} $\rho_{\textrm{st}}$ between predictions and single trial neuronal responses, without averaging across repeats. 
This metric will be computed for both the \sensorium~ and \sensoriump~ tracks to facilitate comparison.
Single trial correlation on the \textit{final test} set will serve as the ultimate ranking score in the \sensoriump~ track to determine competition winners.

\section*{Baseline}
\label{sec:Baseline}
To establish baselines for our competition, we trained simple linear-nonlinear models (LN Baseline) as well as a state-of-the-art convolutional neural network (CNN Baseline, \citet{Lurz2020-ua}) models for both competitions. The resulting model performances can be found in Table~\ref{table:T1}. 
We trained a single model (based on one random seed) for each competition track and model type. We trained each model on the training set of the respective competition track only, not utilizing the pre-training scans to facilitate reproducibility. An example of how to utilize the pre-training scans can be found in our \textit{``starting kit''}. Further improvement in predictive performance can be made by using model ensembles, which can lead to a  performance gain of 5-10\%. For further details on the model architecture and training procedure, refer to the Materials and Methods section.

\section*{Additional Resources}
\label{sec:Resources}

We are releasing our \textit{``starting kit''} that contains the complete code to fit our baseline models as well as explore the full dataset \footnote[2]{\url{https://github.com/sinzlab/sensorium}}. The following material is included:
\begin{itemize}
 	\item Code to download our complete corpus of data and to load it for model fitting.
 	\item A visual guide to the properties of the dataset, to give neuroscientists as well as non-experts an overview of the scale and all properties of the data.
 	\item Code to re-train the baselines, as well as checkpoints to load our pre-trained baselines.
 	\item Ready-to-use code to generate a submission file to enter the competition.
\end{itemize}

By facilitating ease of use, we hope that our competition will be of use to both computational neuroscientists and machine learning practitioners alike to apply their knowledge and model-building skills on an interesting biological problem.

\section*{Discussion}
Here, we introduce the \sensorium~competition to find the best predictive model for the mouse primary visual cortex, with the accompanying release of a large-scale dataset of more than 28,000 neurons across seven mice and their activity in response to natural stimuli. We hope that this competition and public benchmark infrastructure will serve as a catalyst for both computational neuroscientists and machine learning practitioners to advance the field of neuro-predictive modeling. Our broader goal is to continue to populate the underlying benchmark infrastructure with additional future iterations of dataset releases, regular challenges, and additional metrics. 


As is the case for benchmarks in general, by converging in this first iteration on a specific dataset, task, and evaluation metric in order to facilitate constructive comparison, \sensorium~ 2022 also becomes limited to the scope of those choices. In particular, we opted for simplicity for the first competition hosted on our platform in order to appeal to a broader audience across the computational neuroscience and machine learning communities. \textit{A priori}, it is not clear how well the best performing models of this competition would transfer to a broader or more naturalistic setting -- for example, predicting freely moving animals' neural activity to arbitrary naturalistic videos in an ethologically relevant color spectrum.  Having established our benchmarking framework, we plan to alleviate the constraints imposed by the data by extending future challenges in a number of dimensions, such as:

\begin{itemize}
    \item including cortical layers beyond L2/3 and other areas in mouse visual cortex beyond V1
    \item replacing static image stimuli with dynamic movie stimuli in order to better capture the temporal evolution of representation and/or simulation
    \item replacing grayscale visual stimuli with coverage of both UV- and green-sensitive cone photoreceptors
    \item increasing the number of animals and recordings in the test set beyond one per track to emphasize generalization across animals and brain states
    \item moving beyond passive viewing of visual stimuli by incorporating decision making into the stimulus paradigm
    \item including different or multiple sensory domains (e.g., auditory, olfactory, somatosensory, etc) and motor areas
    \item recording neural responses with different techniques (e.g., electrophysiology) that emphasize different population sizes and spatiotemporal resolution
    \item recording neural responses in different animal models, such as non-human primates. 
    \item inverting model architecture to reconstruct visual input from neural responses. 
\end{itemize}

Similarly, selection of appropriate metrics for model performance is also challenging as, despite considerable progress in this domain \citep{Pospisil2020, Schoppe2016, Haefner2008}, there is not yet a consensus metric for evaluating predictive models of the brain.  Ideally, such a metric would appropriately handle different sources of variability present in neural responses and 
have a calibrated upper bound (i.e., where perfect prediction of the stimulus-driven component receives a perfect score).  Metrics such as the fraction of explainable variance explained (FEVE, \citet{Cadena2019}) partially address this issue, but focus on the mean and variance are not applicable in settings when no two trials with identical inputs exist, for example when the input includes uncontrolled behavioral variables.  
On the other hand, likelihood-based metrics can reflect the entire conditional response distribution, but can be sensitive to the choice of noise model and under-emphasize the contribution of the stimulus-driven component. In addition, upper bounds are often challenging to define for these metrics.  

More broadly, a potential concern about predictive modeling in general is that the models with the highest predictive performance generally lack interpretability of how they achieve their performance due to their complexity, which limits scientific insight. However, we think that the converse is equally problematic: a simpler model that is interpretable but poorly predicts neural activity, in particular to natural stimuli, would also be unhelpful. We believe that a fruitful avenue for gaining insight into the brain's functional processing is to iteratively combine both approaches: developing new models to push the state-of-the-art predictive performance and at the same time extracting knowledge by simplifying complex models or by analyzing models post-hoc. For example, \citet{Burg2021} simplified the state-of-the-art model by \citet{Cadena2019} showing that divisive normalization accounts for most but not all of its performance; and \citet{Ustyuzhaninov2022} simplified and analyzed the representations learned by a high-performing complex model revealing a combinatorial code of non-linear computations in mouse V1.

Additionally, high performing predictive models may also benefit computational neuroscientists by serving as digital twins, creating an \textit{in silico} environment in which hypotheses may be developed and refined before returning to the \textit{in vivo} system for validation \citep{Bashivan2019, Ponce2019-yn, Walker2019-oq, Franke2021}.  This approach can not only enable exhaustive experiments that are otherwise technically infeasible \textit{in vivo},
but also limit cost and ethical burden by decreasing the number of animals required to pursue scientific questions. 

Taken together, we believe that predictive models have become an important tool for neuroscience research. In our view, systematically benchmarking and improving these models along with the development of accurate metrics will be of great benefit to neuroscience as a whole. We therefore invite the research community to join the benchmarking effort by participating in the challenge, and by contributing new datasets and metrics to our benchmarking system. We would like to cast the challenge of understanding information processing in the brain as a joint endeavor in which we engage together as a whole community, iteratively re-defining what is the state-of-the-art in predicting neural activity and leveraging models to pursue the question of how the brain makes sense of the world.

\subsection*{Acknowledgments}
KKL is funded by the German Federal Ministry of Education and Research through the Tübingen AI Center (FKZ: 01IS18039A). FHS is supported by the Carl-Zeiss-Stiftung and acknowledges the support of the DFG Cluster of Excellence “Machine Learning – New Perspectives for Science”, EXC 2064/1, project number 390727645. This work was supported by an AWS Machine Learning research award to FHS. MB and SAC were supported by the International Max Planck Research School for Intelligent Systems. This project has received funding from the European Research Council (ERC) under the European Union’s Horizon Europe research and innovation programme (Grant agreement No. 101041669).

This research was supported by National Institutes of Health (NIH) via National Eye Insitute (NEI) grant RO1-EY026927, NEI grant T32-EY002520, National Institute of Mental Health (NIMH) and National Institute of Neurological Disorders and Stroke (NINDS) grant U19-MH114830, and NINDS grant U01-NS113294.  This research was also supported by National Science Foundation (NSF) NeuroNex grant 1707400.  The content is solely the responsibility of the authors and does not necessarily represent the official views of the NIH, NEI, NIMH, NINDS, or NSF.

This research was also supported by the Intelligence Advanced Research Projects Activity (IARPA) via Department of Interior/Interior Business Center (DoI/IBC) contract no. D16PC00003, and with funding from the Defense Advanced Research Projects Agency (DARPA), Contract No. N66001-19-C-4020. The US Government is authorized to reproduce and distribute reprints for governmental purposes notwithstanding any copyright annotation thereon. The views and conclusions contained herein are those of the authors and should not be interpreted as necessarily representing the official policies or endorsements, either expressed or implied, of IARPA, DoI/IBC, DARPA, or the US Government.





\begin{contributions}

\textbf{KFW}: Conceptualization, Methodology, Data curation, Validation, Software, Formal Analysis, Analysis design, Writing - Original Draft \& Editing, Visualization, Project administration;
\textbf{PGF}: Conceptualization, Methodology, Data acquisition, Data curation, Visualization, Writing - Original Draft \& Editing; 
\textbf{MB}:  Conceptualization, Methodology, Software, Analysis design, Writing - Original Draft \& Editing;
\textbf{LP}: Conceptualization, Software, Writing - Original Draft \& Editing; 
\textbf{MFB}: Conceptualization, Software, Writing - Original Draft \& Editing;  
\textbf{CB}: Software, Analysis design ; 
\textbf{SAC}:Conceptualization, Methodology, Software, Analysis design, Writing - Original Draft \& Editing; 
\textbf{ZD}: Data acquisition, Data curation; 
\textbf{KKL}:Methodology, Software, Writing - Original Draft \& Editing;   
\textbf{KP}: Data acquisition, Data curation; 
\textbf{TM}: Data acquisition, Data curation; 
\textbf{SSP}: Data acquisition; 
\textbf{AE}: Conceptualization, Supervision, Writing - Review \& Editing, Funding acquisition;  
\textbf{AST}: Supervision, Data acquisition, Writing - Review \& Editing, Funding acquisition; 
\textbf{FHS}: Conceptualization, Supervision, Writing - Review \& Editing, Funding acquisition;
\end{contributions}


\section*{Materials and Methods}
\label{sec:Methods}
\subsection*{Neurophysiological experiments} 
All procedures were approved by the Institutional Animal Care and Use Committee of Baylor College of Medicine. Seven mice (Mus musculus, 5 females, 2 males) expressing GCaMP6s in excitatory neurons via Slc17a7-Cre and Ai162 transgenic lines (recommended and generously shared by Hongkui Zeng at Allen Institute for Brain Science; JAX stock 023527 and 031562, respectively) were anesthetized and a 4 mm craniotomy was made over the visual cortex of the right hemisphere as described previously \citep{Reimer2014-ry, Froudarakis2014-lx}. 

Mice were head-mounted above a cylindrical treadmill and calcium imaging was performed using Chameleon Ti-Sapphire laser (Coherent) tuned to 920 nm and a large field of view mesoscope \citep{Sofroniew2016-up} equipped with a custom objective (excitation NA 0.6, collection NA 1.0,  21 mm focal length). Laser power after the objective was increased exponentially as a function of depth from the surface according to:

\begin{equation}
    P = P_0 \times e^{(z/L_z)}
\end{equation}

Here P is the laser power used at target depth z, P0 is the power used at the surface (not exceeding 19 mW), and Lz is the depth constant (220 μm).  The greatest laser output of 55 mW was used at approximately 245 μm from the surface, with most scans not requiring more than 40 mW at similar depths.   

The craniotomy window was leveled with regards to the objective with six degrees of freedom.  Pixel-wise responses from an ROI spanning the cortical window (>3000 x 3000 μm, 0.2 px/μm, between 200-220 μm from surface, >3.9 Hz) to drifting bar stimuli were used to generate a sign map for delineating visual areas \citep{Garrett2014-ki}.  Area boundaries on the sign map were manually annotated.  Our target imaging site was a 630 x 630 x 45 um volume (anteroposterior x mediolateral x radial depth) in L2/3, within the boundaries of primary visual cortex (VISp, Supp.Fig.~\ref{fig:SFig_1}).  

The released scans contained 10 planes, with 5 μm interplane distance in depth, and were collected at 7.98 Hz.  Each plane is 630 $\times$ 630 μm (252 $\times$ 252 pixels, 0.4 px/μm).  The most superficial plane in each volume ranged from approximately 190 - 220 μm from the surface.  This high spatial sampling in z results in the nearest plane no more that 2.5 μm from the theoretical best imaging plane for each neuron, but also results in redundant masks for many neurons in adjacent planes.  

Movie of the animal's eye and face was captured throughout the experiment.  A hot mirror (Thorlabs FM02) positioned between the animal's left eye and the stimulus monitor was used to reflect an IR image onto a camera (Genie Nano C1920M, Teledyne Dalsa) without obscuring the visual stimulus.  The position of the mirror and camera were manually calibrated per session and focused on the pupil.  Field of view was manually cropped for each session.  For five pre-training scans, the field of view contained the superior, frontal, and inferior portions of the facial silhouette as well as the left eye in its entirety, ranging from 664-906 pixels height x 1014-1113 pixels width at ~20 Hz.  In two remaining competition scans, the field of view contained only the left eye in its entirety, 282-300 pixels height x 378-444 pixels width at ~20 Hz.  Frame times were time stamped in the behavioral clock for alignment to the stimulus and scan frame times. Video was compressed using Labview’s MJPEG codec with quality constant of 600 and stored in an AVI file.

Light diffusing from the laser during scanning through the pupil was used to capture pupil diameter and eye movements.  A DeepLabCut model \citep{Mathis2018-ti} was trained on 17 manually labeled samples from 11 animals to label each frame of the compressed eye video (intraframe only H.264 compression, CRF:17) with 8 eyelid points and 8 pupil points at cardinal and intercardinal positions.  Pupil points with likelihood >0.9 (all 8 in 92-98\% of frames) were fit with the smallest enclosing circle, and the radius and center of this circle was extracted.  Frames with < 3 pupil points with likelihood >0.9 (<0.3\% frames per scan), or producing a circle fit with outlier > 5.5 standard deviations from the mean in any of the three parameters (center x, center y, radius, <0.5\% frames per scan) were discarded (total <0.6\% frames per scan).  Gaps of <= 2 discarded frames were replaced by linear interpolation (<0.2\% frames per scan).  Trials affected by remaining gaps were discarded (<50 trials per scan, <0.9\%).  

The mouse was head-restrained during imaging but could walk on a treadmill. Rostro-caudal treadmill movement was measured using a rotary optical encoder (Accu-Coder 15T-01SF-2000NV1ROC-F03-S1) with a resolution of 8000 pulses per revolution, and was recorded at 99.8-100.3 Hz in order to extract locomotion velocity. 

\subsection*{Visual stimulation} 

Visual stimuli were presented with Psychtoolbox 3 in MATLAB \citep{Brainard1997-ed, Kleiner2007-ik, Pelli1997-vv} to the left eye with a 31.8 x 56.5 cm (height x width) monitor (ASUS PB258Q) with a resolution of 1080 x 1920 pixels positioned 15 cm away from the eye.  When the monitor is centered on and perpendicular to the surface of the eye at the closest point, this corresponds to a visual angle of 3.8 °/cm at the nearest point and 0.7 °/cm at the most remote corner of the monitor.  As the craniotomy coverslip placement during surgery and the resulting mouse positioning relative to the objective is optimized for imaging quality and stability, uncontrolled variance in animal skull position relative to the washer used for head-mounting was compensated with tailored monitor positioning on a six dimensional monitor arm. The pitch of the monitor was kept in the vertical position for all animals, while the roll was visually matched to the roll of the animal’s head beneath the headbar by the experimenter. In order to optimize the translational monitor position for centered visual cortex stimulation with respect to the imaging field of view, we used a dot stimulus with a bright background (maximum pixel intensity) and a single dark square dot (minimum pixel intensity).  Dot locations were randomly ordered from one of two regimes: a 5 x 8 grid tiling the screen with 20 repetitions of 200 ms presentation at each location, or a 10 x 10 grid tiling a central square (approx 90 degrees width and height) with 10 repetitions of 200-300 ms presentation at each location.  The final monitor position for each animal was chosen in order to center the population receptive field of the scan field ROI on the monitor, with the yaw of the monitor visually matched to be perpendicular to and 15 cm from the nearest surface of the eye at that position.  

\underline{\textit{Natural images}}: Natural images were randomly selected from ImageNet \citep{Russakovsky2015-xi}, converted to gray scale, and cropped to the monitor aspect ratio 16:9. Images were downsampled and stored at $144 \times 256$ ($h \times w$) resolution, and upsampled to $1080 \times 1920$ pixels ($h \times w$) by bilinear interpolation during presentation.  We presented 5100 unique natural images during each pre-training scan, with 5000 of them presented once and the remaining 100 repeated 10 times each (total 6000 trials). The same 100 images were repetitively shown in all scans. In the two competition scans, an additional set of 100 images were shown and repeated 10 times each (total 7000 trials). Each image was presented for 500 ms, preceded by a blank screen lasting for a period uniformly sampled from 300 to 500 ms. Lastly, the two competition scans also included 10 randomly intermixed repeats of 1 minute of dynamic natural movie stimuli that are not part of the training or test sets in this release.

A photodiode (TAOS TSL253) was sealed to the top left corner of the monitor, and the voltage was recorded at 10 MHz and timestamped on the behavior clock (MasterClock PCIe-OSC-HSO-2 card).  Simultaneous measurement with a luminance meter (LS-100 Konica Minolta) perpendicular to and targeting the center of the monitor was used to generate a lookup table for linear interpolation between photodiode voltage and monitor luminance in cd/m\textsuperscript{2} for 16 equidistant values from 0-255, and one baseline value with the monitor unpowered.

At the beginning of each experimental session, we collected photodiode voltage for 52 full-screen pixel values from 0 to 255 for one second trials.  The mean photodiode voltage for each trial $V_{pd}$ was fit as a function of the pixel intensity $V_{in}$ as:

\begin{equation}
     V_{pd} = B + A \times V_{in}^{\gamma}
 \end{equation}

in order to estimate the gamma value of the monitor as 1.74.  All stimuli were shown with no gamma correction.  

During the stimulus presentation, sequence information was encoded in a 3 level signal according to the binary encoding of the flip number assigned in-order.  This signal underwent a sine convolution, allowing for local peak detection to recover the binary signal.  The encoded binary signal was reconstructed for >99\% of the flips.  A linear fit was applied to the trial timestamps in the behavioral and stimulus clocks, and the offset of that fit was applied to the data to align the two clocks, allowing linear interpolation between them. The mean photodiode voltage of the sequence encoding signal at pixel values 0 and 255 was used to estimate the luminance range of the monitor during the stimulus, with minimum values between 0.002 and 0.192 cd/m\textsuperscript{2} and maximum values between 9.1 and 11.8 cd/m\textsuperscript{2} in the released scans.  

\subsection*{Preprocessing of neural responses and behavioral data} 

The full two photon imaging processing pipeline is available at (https://github.com/cajal/pipeline). Raster correction for bidirectional scanning phase row misalignment was performed by iterative greedy search at increasing resolution for the raster phase resulting in the maximum cross-correlation between odd and even rows. Motion correction for global tissue movement was performed by shifting each frame in X and Y to maximize the correlation between the cross-power spectra of a single scan frame and a template image, generated from the Gaussian-smoothed average of the Anscombe transform from the middle 2000 frames of the scan. Neurons were automatically segmented using constrained non-negative matrix factorization, then deconvolved to extract estimates of spiking activity, within the CAIMAN pipeline \citep{Giovannucci2019-gw}.  Cells were further selected by a classifier trained to separate somata versus artifacts based on segmented cell masks, resulting in exclusion of 9.1\% of masks.

Functional and behavioral signals were upsampled to 100 Hz by nearest neighbor interpolation, then all points within a boxcar window from 50 - 550 ms after stimulus onset were averaged to produce the trial value.
The mirror motor coordinates of the centroid of each mask was used to assign anatomical coordinates relative to each other and the experimenter's estimate of the pial surface.  
Notably, centroid positional coordinates do not carry information about position relative to the area boundaries, or relative to neurons in other scans.
Multiple masks corresponding to the intersection of a single neuron's light footprint with multiple imaging planes were detected by iteratively assigning edges between units with anatomical coordinates within 80 μm euclidean distance and with correlation over 0.9 between the full duration deconvolved traces after low pass filtering with a Hann window and downsampling to 1 Hz.  The total number of unique neurons per scan was estimated by the number of unconnected components in the resulting graph per scan.

\subsection*{Model architecture}

Our best performing model of the mouse primary visual cortex is a convolutional neural network (CNN) model, which is split into two parts: the \textit{core} and the \textit{readout} \citep{Sinz2018-sk, Cadena2019, Lurz2020-ua, Franke2021}. First, the core computes latent features from the inputs. Then, the readout is learned per neuron and maps the output of the rich core feature space to the neuronal responses via regularized regression. This architecture reflects the inductive bias of a shared feature space across neurons. In this view, each neuron represents a linear combination of features realized at a certain position in the visual field, known as as the receptive field.
\label{sub:networks}

\paragraph{Representation/Core}
We based our work on the models of \citet{Lurz2020-ua} and \citet{Franke2021}, which are able to predict the responses of a large population of mouse V1 neurons with high accuracy. In brief, we modeled the core as a four-layer CNN, with 64 feature channels per layer. In each layer, a 2d-convolution is followed by batch normalization and an ELU nonlinearity \citep{Ioffe2015-or,Clevert2015-tc}. Except for the first layer, all convolutions are depth-separable \citep{Chollet2017}. The latent features that the core computed for a given input are then forwarded to the readout.

\paragraph{{Readout}}
To get the scalar neuronal firing rate for each neuron, we computed a linear regression between the core output tensor of dimensions $\mathbf{x}\in \mathbb{R}^{w \times h \times c}$ (\textbf{w}idth, \textbf{h}eight, \textbf{c}hannels) and the linear weight tensor $\mathbf{w} \in \mathbb{R}^{c \times w \times h}$, followed by an ELU offset by one (ELU+1), to keep the response positive. We made use of the recently proposed Gaussian readout \citep{Lurz2020-ua}, which simplifies the regression problem considerably.
The Gaussian readout learns the parameters of a 2D Gaussian distribution $\mathcal{N}(\mu_n, \Sigma_n)$. The mean $\mu_n$ in the readout feature space thus represents the center of a neuron's receptive field in image space, whereas $\Sigma_n$ refers to the uncertainty of the receptive field position. During training, a location of height and width in the core output tensor in each training step is sampled, for every image and neuron. Given a large enough initial $\Sigma_n$ to ensure gradient flow, the uncertainty about the readout location $\Sigma_n$ is decreasing during training, showing that the estimates of the mean location $\mu_n$ becomes more and more reliable.  At inference time (i.e. when evaluating our model), we set the readout to be deterministic and to use the fixed position $\mu_n$. In parallel to learning the position, we learned the weights of the weight tensor of the linear regression of size $c$ per neuron. To learn the positions $\mu_n$, we made use of the retinotopic organization of V1 by coupling the recorded cortical 2d-coordinates $\mathbf{p}_n\in \mathbb R^2$ of each neuron with the estimation of the receptive field position $\mu_n$ of the readout. We achieved this by learning the common function $\mu_n=f(\mathbf{p}_n)$, a randomly initialized linear fully connected MLP of size 2-30-2, shared by all neurons.

\paragraph{{Shifter network}}
We employed a free viewing paradigm when presenting the visual stimuli to the head-fixed mice. Thus, the RF positions of the neurons with respect to the presented images had considerable trial to trial variability following any eye movements. We informed our model of the trial dependent shift of the neurons receptive fields due to eye movement by shifting $\mu_n$, the model neuron's receptive field center, using the estimated eye position (see section Neurophysiological experiments above for details of estimating the pupil center). We passed the estimated pupil center through a MLP, called shifter network, a three layer fully connected network with $n=5$ hidden features, followed by a $tanh$ nonlinearity, that calculates the shift in $\Delta$x and $\Delta$y of the neurons receptive field in each trial. We then added this shift to the $\mu_n$ of each neuron.

\paragraph{{Input of behavioral parameters}}
During each presentation of a natural image, the pupil size, instantaneous change of pupil size, and the running speed of the mouse was recorded. For the \sensoriump~ track, we have used these behavioral parameters to improve the model's predictivity. Because these behavioral parameters have nonlinear modulatory effects, we decided to appended them as separate images to the input images as new channels \citep{Franke2021}, such that each new channel simply consisted of the scalar for the respective behavioral parameter recorded in a particular trial, transformed into stimulus dimension. This enabled the model to predict neural responses as a function of both visual input and behavior. 

\subsection*{Model training}
We first split the unique training images into the training and validation set, using a split of 90\% to 10\%, respectively. Subsequently, we isotropically downsampled all images to a resolution of $36 \times 64$ px ($h \times w$), resulting in a resolution of 0.53 pixels per degree visual angle. Furthermore, we normalized Input images as well behavioral traces and standardized the target neuronal activities, using the statistics of the training trials of each recording. Then, we trained our networks with the training set by minimizing the Poisson loss $\frac{1}{m}\sum_{i=1}^{m}\big(\hat{r}^{(i)}-r^{(i)}\log{\hat{r}^{(i)}}\big)$, where $m$ denotes the number of neurons, $\hat r$ the predicted neuronal response and $r$ the observed response. After each epoch, i.e. full pass through the training set, we calculated the correlation between predicted and measured neuronal responses on the validation set and averaged it across all neurons. If the correlation failed to increase during for five consecutive epochs, we stopped the training and restored the model to its state after the best performing epoch. Then, we either decreased the learning rate by a factor of $0.3$ or stopped training altogether, if the number of learning-rate decay steps was reached (n=4 decay steps). We optimized the network's parameters using the Adam optimizer \citep{kingma2014adam}. 
Furthermore, we performed an exhaustive hyper-parameter selection using Bayesian search on a held-out dataset. All parameters and hyper-parameters regarding model architecture and training procedure can be found in our \code{sensorium} repository (see Code Availability).

\subsection*{LN model}
Our LN model has the same architecture as our CNN model, but with all non-linearities removed except the final ELU+1 nonlinearity, thus turning the CNN model effectively into a fully linear model followed by a single output non-linearity. We performed a search for the best architectural parameters on a held-out dataset and arrived at a model with three layers, and otherwise unchanged parameters as compared to the CNN model. The addition of the shifter network, behavioral parameters, as well as the whole training procedure, was identical to the CNN model.

\subsection*{Metrics}
\label{sec:Metrics}
We chose to evaluate the models via two metrics that are commonly used in neuroscience but differ in their properties and focus: Correlation and Fraction of Explainable Variance Explained (FEVE).\\
Since \textbf{correlation} is invariant to shift and scale of the predictions, it does not reward a correct prediction of the absolute value of the neural response but rather the neuron's relative response changes. It is bound to $[-1, 1]$ and thus easily interpretable. However, without accounting for the unexplainable noise in neural responses, the upper bound of $1$ cannot be reached, which can be misleading. \\
\textbf{FEVE}, on the other hand, does correct for the unexplainable noise and provides provides an estimate for the upper bound of stimulus-driven activity. This measure is equivalent to variance explained ($R^2$) corrected for observation noise. However, since its bounds are $[-\infty, 1]$, models can be arbitrarily bad. If the model predicts the constant mean of the target responses, FEVE will evaluate to $0$. In the case of single-trial model evaluation in the \sensoriump~ track, FEVE is not possible to compute, as it requires stimulus trial repeats to estimate the unexplainable variability.\\
Both metrics are restricted in that they are point estimates, and as such do not measure the correctness of the entire probability distribution that the model has learned. Instead, they only measure the correctness of the first moment of this distribution, i.e. the mean. 

\textbf{Correlation to Average}
Given a neuron's response $r_{ij}$ to image $i$ and repeat $j$ and predictions $o_{i}$, the correlation is computed between the predicted responses and the average neuronal response $\overline{r}_i$ to the $i^{th}$ test image (averaged across repeated presentations of the same stimulus):
 
 \begin{equation}
     \rho_{\textrm{av}} = \textrm{corr}(\mathbf{r}_{\textrm{av}}, \mathbf{o}_{\textrm{av}}) = \frac{\sum_{i}(\bar{r}_{i} - \bar{r})(o_{i} - \bar{o})}{\sqrt{\sum_{i}(\bar{r}_{i} - \bar{r})^2 \sum_{i}(o_{i} - \bar{o})^2}},
 \end{equation}

 where $\bar{r}_{i} = \frac{1}{J} \sum_{j=1}^{J} r_{ij}$ is the average response across $J$ repeats, $\bar{r}$ is the average response across all repeats and images, and $\bar{o}$ is the average prediction across all repeats and images. 

 \textbf{Fraction Explainable Variance Explained}
 For $N=I\cdot J$ is the total number of trials from $I$ images and $J$ repeats per image:
 
 \begin{equation}
     \textrm{FEVE} = 1 - \frac{\frac{1}{N} \sum_{i, j} (r_{ij} - o_{i})^2 - \sigma_{\varepsilon}^2}{\textrm{Var}[\mathbf{r}] - \sigma_{\varepsilon}^2},
 \end{equation}

 where $\textrm{Var}[\mathbf{r}]$ is the total response variance computed across all $N$ trials and $\sigma_{\varepsilon}^2 = \mathbb{E}_i[\textrm{Var}_j[\mathbf{r}|x]]$ is the observation noise variance computed as the average variance across responses to repeated presentations of the stimulus $ x $. As the bounds of FEVE are $[-\infty, 1]$, neurons with low explainable variance can have a substantially negative FEVE value and thus bias the population average. We therefore restrict the calculation of the FEVE to all units with an explainable variance larger than 15\%, in line with previous literature \citep{Cadena2019}.

 \textbf{Single Trial Correlation}
 To measure how well models account for trial-to-trial variations we compute correlation $\rho_{\textrm{st}}$ between predictions and single trial neuronal responses, without averaging across repeats. 
 
 \begin{equation}
     \rho_{\textrm{st}} = \textrm{corr}(\mathbf{r}_{\textrm{st}}, \mathbf{o}_{\textrm{st}}) = \frac{\sum_{i,j}(r_{ij} - \bar{r})(o_{ij} - \bar{o})}{\sqrt{\sum_{i,j}(r_{ij} - \bar{r})^2 \sum_{i,j}(o_{ij} - \bar{o})^2}},
 \end{equation}

 where $\bar{r}$ is the average response across all repeats and images, and $\bar{o}$ is the average prediction across all repeats and images.

\subsection*{Code and data availability}
Our competition website can be reached under \url{http://sensorium2022.net/}. There, the complete datasets are available for download.
Our coding framework uses general tools like PyTorch, Numpy, scikit-image, matplotlib, seaborn, DataJoint, Jupyter, Docker, CAIMAN, DeepLabCut, Psychtoolbox, Scanimage, and Kubernetes. We also used the following custom libraries and code: \code{neuralpredictors} (\url{https://github.com/sinzlab/neuralpredictors}) for torch-based custom functions for model implementation, \code{nnfabrik} (\url{https://github.com/sinzlab/nnfabrik}) for automatic model training pipelines, and \code{sensorium} for utilities (\url{https://github.com/sinzlab/sensorium}).



\section*{References}
\bibliographystyle{apa-good}
\bibliography{references}

\begin{thebibliography}{60}
\expandafter\ifx\csname natexlab\endcsname\relax\def\natexlab#1{#1}\fi
\expandafter\ifx\csname url\endcsname\relax
  \def\url#1{{\tt #1}}\fi
\expandafter\ifx\csname urlprefix\endcsname\relax\def\urlprefix{URL }\fi

\bibitem[{Adelson \& Bergen(1985)}]{Adelson1985-re}
Adelson, E.~H., \& Bergen, J.~R. (1985).
\newblock Spatiotemporal energy models for the perception of motion.
\newblock {\em J. Opt. Soc. Am.\/}, {\em 2\/}(2), 284--299.

\bibitem[{Antol{\'\i}k et~al.(2016)Antol{\'\i}k, Hofer, Bednar, \&
  Mrsic-flogel}]{Antolik2016}
Antol{\'\i}k, J., Hofer, S.~B., Bednar, J.~A., \& Mrsic-flogel, T.~D. (2016).
\newblock Model constrained by visual hierarchy improves prediction of neural
  responses to natural scenes.
\newblock {\em PLoS Comput. Biol.\/}, (pp. 1--22).

\bibitem[{Bashiri et~al.(2021)Bashiri, Walker, Lurz, Jagadish, Muhammad, Ding,
  Ding, Tolias, \& Sinz}]{Bashiri2021-or}
Bashiri, M., Walker, E., Lurz, K.-K., Jagadish, A., Muhammad, T., Ding, Z.,
  Ding, Z., Tolias, A., \& Sinz, F. (2021).
\newblock A flow-based latent state generative model of neural population
  responses to natural images.
\newblock {\em Adv. Neural Inf. Process. Syst.\/}, {\em 34\/}.

\bibitem[{Bashivan et~al.(2019)Bashivan, Kar, \& DiCarlo}]{Bashivan2019}
Bashivan, P., Kar, K., \& DiCarlo, J.~J. (2019).
\newblock {Neural population control via deep image synthesis.}
\newblock {\em Science (New York, N.Y.)\/}, {\em 364\/}(6439).

\bibitem[{Batty et~al.(2016)Batty, Merel, Brackbill, Heitman, Sher, Litke,
  Chichilnisky, \& Paninski}]{Batty2016}
Batty, E., Merel, J., Brackbill, N., Heitman, A., Sher, A., Litke, A.,
  Chichilnisky, E.~J., \& Paninski, L. (2016).
\newblock Multilayer network models of primate retinal ganglion cells.

\bibitem[{Brainard(1997)}]{Brainard1997-ed}
Brainard, D.~H. (1997).
\newblock The psychophysics toolbox.
\newblock {\em Spat. Vis.\/}, {\em 10\/}(4), 433--436.

\bibitem[{Burg et~al.(2021)Burg, Cadena, Denfield, Walker, Tolias, Bethge, \&
  Ecker}]{Burg2021}
Burg, M.~F., Cadena, S.~A., Denfield, G.~H., Walker, E.~Y., Tolias, A.~S.,
  Bethge, M., \& Ecker, A.~S. (2021).
\newblock Learning divisive normalization in primary visual cortex.
\newblock {\em PLOS Computational Biology\/}, {\em 17\/}(6), e1009028.

\bibitem[{Cadena et~al.(2019)Cadena, Denfield, Walker, Gatys, Tolias, Bethge,
  \& Ecker}]{Cadena2019}
Cadena, S.~A., Denfield, G.~H., Walker, E.~Y., Gatys, L.~A., Tolias, A.~S.,
  Bethge, M., \& Ecker, A.~S. (2019).
\newblock Deep convolutional models improve predictions of macaque v1 responses
  to natural images.
\newblock {\em {PLOS} Computational Biology\/}, {\em 15\/}(4), e1006897.

\bibitem[{Cadieu et~al.(2014)Cadieu, Hong, Yamins, Pinto, Ardila, Solomon,
  Majaj, \& DiCarlo}]{Cadieu2014}
Cadieu, C.~F., Hong, H., Yamins, D. L.~K., Pinto, N., Ardila, D., Solomon,
  E.~A., Majaj, N.~J., \& DiCarlo, J.~J. (2014).
\newblock Deep neural networks rival the representation of primate {IT} cortex
  for core visual object recognition.
\newblock {\em PLoS Comput. Biol.\/}, {\em 10\/}(12), e1003963.

\bibitem[{Carandini et~al.(2005)Carandini, Demb, Mante, Tolhurst, Dan,
  Olshausen, Gallant, \& Rust}]{carandiniWeKnowWhat2005}
Carandini, M., Demb, J.~B., Mante, V., Tolhurst, D.~J., Dan, Y., Olshausen,
  B.~A., Gallant, J.~L., \& Rust, N.~C. (2005).
\newblock Do we know what the early visual system does?
\newblock {\em J. Neurosci.\/}, {\em 25\/}(46), 10577--10597.

\bibitem[{Chollet(2017)}]{Chollet2017}
Chollet, F. (2017).
\newblock {Xception: Deep learning with depthwise separable convolutions}.
\newblock In {\em Proceedings - 30th IEEE Conference on Computer Vision and
  Pattern Recognition, CVPR 2017\/}.

\bibitem[{Cichy et~al.(2021)Cichy, Dwivedi, Lahner, Lascelles, Iamshchinina,
  Graumann, Andonian, Murty, Kay, Roig, \& Oliva}]{Algonauts2021}
Cichy, R.~M., Dwivedi, K., Lahner, B., Lascelles, A., Iamshchinina, P.,
  Graumann, M., Andonian, A., Murty, N. A.~R., Kay, K., Roig, G., \& Oliva, A.
  (2021).
\newblock The algonauts project 2021 challenge: How the human brain makes sense
  of a world in motion.
\newline\urlprefix\url{https://arxiv.org/abs/2104.13714}

\bibitem[{Clevert et~al.(2015)Clevert, Unterthiner, \&
  Hochreiter}]{Clevert2015-tc}
Clevert, D.-A., Unterthiner, T., \& Hochreiter, S. (2015).
\newblock Fast and accurate deep network learning by exponential linear units
  ({ELUs}).

\bibitem[{Cowley \& Pillow(2020)}]{Cowley2020neurips}
Cowley, B., \& Pillow, J. (2020).
\newblock High-contrast "gaudy" images improve the training of deep neural
  network models of visual cortex.
\newblock In H.~Larochelle, M.~Ranzato, R.~Hadsell, M.~Balcan, \& H.~Lin (Eds.)
  {\em Advances in Neural Information Processing Systems 33\/}, (pp.
  21591--21603). Curran Associates, Inc.

\bibitem[{de~Vries et~al.(2019)de~Vries, Lecoq, Buice, Groblewski, Ocker,
  Oliver, Feng, Cain, Ledochowitsch, Millman, Roll, Garrett, Keenan, Kuan,
  Mihalas, Olsen, Thompson, Wakeman, Waters, Williams, Barber, Berbesque,
  Blanchard, Bowles, Caldejon, Casal, Cho, Cross, Dang, Dolbeare, Edwards,
  Galbraith, Gaudreault, Gilbert, Griffin, Hargrave, Howard, Huang, Jewell,
  Keller, Knoblich, Larkin, Larsen, Lau, Lee, Lee, Leon, Li, Long, Luviano,
  Mace, Nguyen, Perkins, Robertson, Seid, Shea-Brown, Shi, Sjoquist,
  Slaughterbeck, Sullivan, Valenza, White, Williford, Witten, Zhuang, Zeng,
  Farrell, Ng, Bernard, Phillips, Reid, \& Koch}]{deVries2019}
de~Vries, S. E.~J., Lecoq, J.~A., Buice, M.~A., Groblewski, P.~A., Ocker,
  G.~K., Oliver, M., Feng, D., Cain, N., Ledochowitsch, P., Millman, D., Roll,
  K., Garrett, M., Keenan, T., Kuan, L., Mihalas, S., Olsen, S., Thompson, C.,
  Wakeman, W., Waters, J., Williams, D., Barber, C., Berbesque, N., Blanchard,
  B., Bowles, N., Caldejon, S.~D., Casal, L., Cho, A., Cross, S., Dang, C.,
  Dolbeare, T., Edwards, M., Galbraith, J., Gaudreault, N., Gilbert, T.~L.,
  Griffin, F., Hargrave, P., Howard, R., Huang, L., Jewell, S., Keller, N.,
  Knoblich, U., Larkin, J.~D., Larsen, R., Lau, C., Lee, E., Lee, F., Leon, A.,
  Li, L., Long, F., Luviano, J., Mace, K., Nguyen, T., Perkins, J., Robertson,
  M., Seid, S., Shea-Brown, E., Shi, J., Sjoquist, N., Slaughterbeck, C.,
  Sullivan, D., Valenza, R., White, C., Williford, A., Witten, D.~M., Zhuang,
  J., Zeng, H., Farrell, C., Ng, L., Bernard, A., Phillips, J.~W., Reid, R.~C.,
  \& Koch, C. (2019).
\newblock A large-scale standardized physiological survey reveals functional
  organization of the mouse visual cortex.
\newblock {\em Nature Neuroscience\/}, {\em 23\/}(1), 138--151.
\newline\urlprefix\url{https://doi.org/10.1038/s41593-019-0550-9}

\bibitem[{Ecker et~al.(2018)Ecker, Sinz, Froudarakis, Fahey, Cadena, Walker,
  Cobos, Reimer, Tolias, \& Bethge}]{Ecker2018}
Ecker, A.~S., Sinz, F.~H., Froudarakis, E., Fahey, P.~G., Cadena, S.~A.,
  Walker, E.~Y., Cobos, E., Reimer, J., Tolias, A.~S., \& Bethge, M. (2018).
\newblock A rotation-equivariant convolutional neural network model of primary
  visual cortex.

\bibitem[{Franke et~al.(2021)Franke, Willeke, Ponder, Galdamez, Muhammad,
  Patel, Froudarakis, Reimer, Sinz, \& Tolias}]{Franke2021}
Franke, K., Willeke, K.~F., Ponder, K., Galdamez, M., Muhammad, T., Patel, S.,
  Froudarakis, E., Reimer, J., Sinz, F., \& Tolias, A.~S. (2021).
\newblock Behavioral state tunes mouse vision to ethological features through
  pupil dilation.
\newblock {\em bioRxiv\/}.
\newline\urlprefix\url{https://www.biorxiv.org/content/early/2021/10/13/2021.09.03.458870}

\bibitem[{Froudarakis et~al.(2014)Froudarakis, Berens, Ecker, Cotton, Sinz,
  Yatsenko, Saggau, Bethge, \& Tolias}]{Froudarakis2014-lx}
Froudarakis, E., Berens, P., Ecker, A.~S., Cotton, R.~J., Sinz, F.~H.,
  Yatsenko, D., Saggau, P., Bethge, M., \& Tolias, A.~S. (2014).
\newblock Population code in mouse {V1} facilitates readout of natural scenes
  through increased sparseness.
\newblock {\em Nat. Neurosci.\/}, {\em 17\/}(6), 851--857.

\bibitem[{Garrett et~al.(2014)Garrett, Nauhaus, Marshel, \&
  Callaway}]{Garrett2014-ki}
Garrett, M.~E., Nauhaus, I., Marshel, J.~H., \& Callaway, E.~M. (2014).
\newblock Topography and areal organization of mouse visual cortex.
\newblock {\em J. Neurosci.\/}, {\em 34\/}(37), 12587--12600.

\bibitem[{Giovannucci et~al.(2019)Giovannucci, Friedrich, Gunn, Kalfon, Brown,
  Koay, Taxidis, Najafi, Gauthier, Zhou, Khakh, Tank, Chklovskii, \&
  Pnevmatikakis}]{Giovannucci2019-gw}
Giovannucci, A., Friedrich, J., Gunn, P., Kalfon, J., Brown, B.~L., Koay,
  S.~A., Taxidis, J., Najafi, F., Gauthier, J.~L., Zhou, P., Khakh, B.~S.,
  Tank, D.~W., Chklovskii, D.~B., \& Pnevmatikakis, E.~A. (2019).
\newblock {CaImAn}: An open source tool for scalable calcium imaging data
  analysis.
\newblock {\em Elife\/}, {\em 8\/}, e38173.

\bibitem[{Haefner \& Cumming(2008)}]{Haefner2008}
Haefner, R., \& Cumming, B. (2008).
\newblock An improved estimator of variance explained in the presence of noise.
\newblock In D.~Koller, D.~Schuurmans, Y.~Bengio, \& L.~Bottou (Eds.) {\em
  Advances in Neural Information Processing Systems\/}, vol.~21. Curran
  Associates, Inc.
\newline\urlprefix\url{https://proceedings.neurips.cc/paper/2008/file/2ab56412b1163ee131e1246da0955bd1-Paper.pdf}

\bibitem[{Heeger(1992{\natexlab{a}})}]{Heeger1992-ig}
Heeger, D.~J. (1992{\natexlab{a}}).
\newblock Half-squaring in responses of cat striate cells.
\newblock {\em Vis. Neurosci.\/}, {\em 9\/}(5), 427--443.

\bibitem[{Heeger(1992{\natexlab{b}})}]{Heeger1992-xx}
Heeger, D.~J. (1992{\natexlab{b}}).
\newblock Normalization of cell responses in cat striate cortex.
\newblock {\em Vis. Neurosci.\/}, {\em 9\/}(2), 181--197.

\bibitem[{Ioffe \& Szegedy(2015)}]{Ioffe2015-or}
Ioffe, S., \& Szegedy, C. (2015).
\newblock Batch normalization: accelerating deep network training by reducing
  internal covariate shift.
\newblock In {\em Proceedings of the 32nd International Conference on Machine
  Learning\/}, ICML'15, (pp. 448--456). JMLR.org.

\bibitem[{Jones \& Palmer(1987)}]{Jones1987-sn}
Jones, J.~P., \& Palmer, L.~A. (1987).
\newblock The two-dimensional spatial structure of simple receptive fields in
  cat striate cortex.
\newblock {\em J. Neurophysiol.\/}, {\em 58\/}(6), 1187--1211.

\bibitem[{Kindel et~al.(2017)Kindel, Christensen, \& Zylberberg}]{Kindel2017}
Kindel, W.~F., Christensen, E.~D., \& Zylberberg, J. (2017).
\newblock Using deep learning to reveal the neural code for images in primary
  visual cortex.

\bibitem[{Kingma \& Ba(2015)}]{kingma2014adam}
Kingma, D.~P., \& Ba, J. (2015).
\newblock Adam: {A} method for stochastic optimization.
\newblock In Y.~Bengio, \& Y.~LeCun (Eds.) {\em 3rd International Conference on
  Learning Representations, {ICLR} 2015, San Diego, CA, USA, May 7-9, 2015,
  Conference Track Proceedings\/}.

\bibitem[{Kleiner et~al.(2007)Kleiner, Brainard, Pelli, Ingling, \&
  Broussard}]{Kleiner2007-ik}
Kleiner, M.~B., Brainard, D.~H., Pelli, D.~G., Ingling, A., \& Broussard, C.
  (2007).
\newblock What's new in psychtoolbox-3.
\newblock {\em Perception\/}, {\em 36\/}, 1--16.

\bibitem[{Klindt et~al.(2017)Klindt, Ecker, Euler, \& Bethge}]{Klindt2017}
Klindt, D.~A., Ecker, A.~S., Euler, T., \& Bethge, M. (2017).
\newblock Neural system identification for large populations separating
  ``what'' and ``where''.
\newblock In {\em Advances in Neural Information Processing Systems\/}, (pp.
  4--6).

\bibitem[{Lau et~al.(2002)Lau, Stanley, \& Dan}]{Lau2002}
Lau, B., Stanley, G.~B., \& Dan, Y. (2002).
\newblock Computational subunits of visual cortical neurons revealed by
  artificial neural networks.
\newblock {\em Proceedings of the National Academy of Sciences\/}, {\em
  99\/}(13), 8974--8979.

\bibitem[{Lehky et~al.(1992)Lehky, Sejnowski, \& Desimone}]{Lehky1992}
Lehky, S., Sejnowski, T., \& Desimone, R. (1992).
\newblock Predicting responses of nonlinear neurons in monkey striate cortex to
  complex patterns.
\newblock {\em The Journal of Neuroscience\/}, {\em 12\/}(9), 3568--3581.
\newline\urlprefix\url{https://doi.org/10.1523/jneurosci.12-09-03568.1992}

\bibitem[{Li et~al.(2019)Li, Brendel, Walker, Cobos, Muhammad, Reimer, Bethge,
  Sinz, Pitkow, \& Tolias}]{Li2019-il}
Li, Z., Brendel, W., Walker, E.~Y., Cobos, E., Muhammad, T., Reimer, J.,
  Bethge, M., Sinz, F.~H., Pitkow, X., \& Tolias, A.~S. (2019).
\newblock Learning from brains how to regularize machines.

\bibitem[{Li et~al.(2022)Li, Caro, Rusak, Brendel, Bethge, Anselmi, Patel,
  Tolias, \& Pitkow}]{Li2022-ot}
Li, Z., Caro, J.~O., Rusak, E., Brendel, W., Bethge, M., Anselmi, F., Patel,
  A.~B., Tolias, A.~S., \& Pitkow, X. (2022).
\newblock Robust deep learning object recognition models rely on low frequency
  information in natural images.

\bibitem[{Lurz et~al.(2020)Lurz, Bashiri, Willeke, Jagadish, Wang, Walker,
  Cadena, Muhammad, Cobos, Tolias, Ecker, \& Sinz}]{Lurz2020-ua}
Lurz, K.-K., Bashiri, M., Willeke, K., Jagadish, A.~K., Wang, E., Walker,
  E.~Y., Cadena, S.~A., Muhammad, T., Cobos, E., Tolias, A.~S., Ecker, A.~S.,
  \& Sinz, F.~H. (2020).
\newblock Generalization in data-driven models of primary visual cortex.
\newblock In {\em Proceedings of the International Conference for Learning
  Representations ({ICLR})\/}, (p. 2020.10.05.326256).

\bibitem[{Mathis et~al.(2018)Mathis, Mamidanna, Cury, Abe, Murthy, Mathis, \&
  Bethge}]{Mathis2018-ti}
Mathis, A., Mamidanna, P., Cury, K.~M., Abe, T., Murthy, V.~N., Mathis, M.~W.,
  \& Bethge, M. (2018).
\newblock {DeepLabCut}: markerless pose estimation of user-defined body parts
  with deep learning.
\newblock {\em Nat. Neurosci.\/}, {\em 21\/}(9), 1281--1289.

\bibitem[{McIntosh et~al.(2016)McIntosh, Maheswaranathan, Nayebi, Ganguli, \&
  Baccus}]{McIntosh2016}
McIntosh, L.~T., Maheswaranathan, N., Nayebi, A., Ganguli, S., \& Baccus, S.~A.
  (2016).
\newblock Deep learning models of the retinal response to natural scenes.
\newblock {\em Adv. Neural Inf. Process. Syst.\/}, {\em 29\/}(Nips),
  1369--1377.

\bibitem[{Niell \& Stryker(2010)}]{niell2010modulation}
Niell, C.~M., \& Stryker, M.~P. (2010).
\newblock Modulation of visual responses by behavioral state in mouse visual
  cortex.
\newblock {\em Neuron\/}, {\em 65\/}(4), 472--479.

\bibitem[{Pei et~al.(2021)Pei, Ye, Zoltowski, Wu, Chowdhury, Sohn, O'Doherty,
  Shenoy, Kaufman, Churchland, Jazayeri, Miller, Pillow, Park, Dyer, \&
  Pandarinath}]{latentsbench}
Pei, F., Ye, J., Zoltowski, D., Wu, A., Chowdhury, R.~H., Sohn, H., O'Doherty,
  J.~E., Shenoy, K.~V., Kaufman, M.~T., Churchland, M., Jazayeri, M., Miller,
  L.~E., Pillow, J., Park, I.~M., Dyer, E.~L., \& Pandarinath, C. (2021).
\newblock Neural latents benchmark '21: Evaluating latent variable models of
  neural population activity.
\newline\urlprefix\url{https://arxiv.org/abs/2109.04463}

\bibitem[{Pelli(1997)}]{Pelli1997-vv}
Pelli, D.~G. (1997).
\newblock The {VideoToolbox} software for visual psychophysics: transforming
  numbers into movies.
\newblock {\em Spat. Vis.\/}, {\em 10\/}(4), 437--442.

\bibitem[{Ponce et~al.(2019)Ponce, Xiao, Schade, Hartmann, Kreiman, \&
  Livingstone}]{Ponce2019-yn}
Ponce, C.~R., Xiao, W., Schade, P.~F., Hartmann, T.~S., Kreiman, G., \&
  Livingstone, M.~S. (2019).
\newblock Evolving images for visual neurons using a deep generative network
  reveals coding principles and neuronal preferences.
\newblock {\em Cell\/}, {\em 177\/}(4), 999--1009.e10.

\bibitem[{Pospisil \& Bair(2020)}]{Pospisil2020}
Pospisil, D.~A., \& Bair, W. (2020).
\newblock The unbiased estimation of the fraction of variance explained by a
  model.

\bibitem[{Prenger et~al.(2004)Prenger, Wu, David, \& Gallant}]{Prenger2004}
Prenger, R., Wu, M. C.-K., David, S.~V., \& Gallant, J.~L. (2004).
\newblock Nonlinear {V1} responses to natural scenes revealed by neural network
  analysis.
\newblock {\em Neural Netw.\/}, {\em 17\/}(5-6), 663--679.

\bibitem[{Reimer et~al.(2014)Reimer, Froudarakis, Cadwell, Yatsenko, Denfield,
  \& Tolias}]{Reimer2014-ry}
Reimer, J., Froudarakis, E., Cadwell, C.~R., Yatsenko, D., Denfield, G.~H., \&
  Tolias, A.~S. (2014).
\newblock Pupil fluctuations track fast switching of cortical states during
  quiet wakefulness.
\newblock {\em Neuron\/}, {\em 84\/}(2), 355--362.

\bibitem[{Russakovsky et~al.(2015)Russakovsky, Deng, Su, Krause, Satheesh, Ma,
  Huang, Karpathy, Khosla, Bernstein, Berg, \& Fei-Fei}]{Russakovsky2015-xi}
Russakovsky, O., Deng, J., Su, H., Krause, J., Satheesh, S., Ma, S., Huang, Z.,
  Karpathy, A., Khosla, A., Bernstein, M., Berg, A.~C., \& Fei-Fei, L. (2015).
\newblock {ImageNet} large scale visual recognition challenge.
\newblock {\em Int. J. Comput. Vis.\/}, {\em 115\/}(3), 211--252.

\bibitem[{Rust et~al.(2005)Rust, Schwartz, Movshon, \& Simoncelli}]{Rust2005}
Rust, N.~C., Schwartz, O., Movshon, J.~A., \& Simoncelli, E.~P. (2005).
\newblock Spatiotemporal elements of macaque v1 receptive fields.
\newblock {\em Neuron\/}, {\em 46\/}(6), 945--956.

\bibitem[{Safarani et~al.(2021)Safarani, Nix, Willeke, Cadena, Restivo,
  Denfield, Tolias, \& Sinz}]{Safarani2021-yy}
Safarani, S., Nix, A., Willeke, K., Cadena, S., Restivo, K., Denfield, G.,
  Tolias, A., \& Sinz, F. (2021).
\newblock Towards robust vision by multi-task learning on monkey visual cortex.
\newblock {\em Adv. Neural Inf. Process. Syst.\/}, {\em 34\/}, 739--751.

\bibitem[{Schoppe et~al.(2016)Schoppe, Harper, Willmore, King, \&
  Schnupp}]{Schoppe2016}
Schoppe, O., Harper, N.~S., Willmore, B. D.~B., King, A.~J., \& Schnupp, J.
  W.~H. (2016).
\newblock Measuring the performance of neural models.
\newblock {\em Frontiers in Computational Neuroscience\/}, {\em 10\/}.
\newline\urlprefix\url{https://doi.org/10.3389/fncom.2016.00010}

\bibitem[{Schrimpf et~al.(2018)Schrimpf, Kubilius, Hong, Majaj, Rajalingham,
  Issa, Kar, Bashivan, Prescott-Roy, Geiger, Schmidt, Yamins, \&
  DiCarlo}]{Schrimpf2018}
Schrimpf, M., Kubilius, J., Hong, H., Majaj, N.~J., Rajalingham, R., Issa,
  E.~B., Kar, K., Bashivan, P., Prescott-Roy, J., Geiger, F., Schmidt, K.,
  Yamins, D. L.~K., \& DiCarlo, J.~J. (2018).
\newblock Brain-score: Which artificial neural network for object recognition
  is most brain-like?
\newline\urlprefix\url{https://doi.org/10.1101/407007}

\bibitem[{Schwartz et~al.(2006)Schwartz, Pillow, Rust, \&
  Simoncelli}]{Schwartz2006-ji}
Schwartz, O., Pillow, J.~W., Rust, N.~C., \& Simoncelli, E.~P. (2006).
\newblock Spike-triggered neural characterization.
\newblock {\em J. Vis.\/}, {\em 6\/}(4), 484--507.

\bibitem[{Sinz et~al.(2018)Sinz, Ecker, Fahey, Walker, Cobos, Froudarakis,
  Yatsenko, Pitkow, Reimer, \& Tolias}]{Sinz2018-sk}
Sinz, F., Ecker, A.~S., Fahey, P., Walker, E., Cobos, E., Froudarakis, E.,
  Yatsenko, D., Pitkow, X., Reimer, J., \& Tolias, A. (2018).
\newblock Stimulus domain transfer in recurrent models for large scale cortical
  population prediction on video.
\newblock In {\em Advances in Neural Information Processing Systems 31\/}.

\bibitem[{Sinz et~al.(2019)Sinz, Pitkow, Reimer, Bethge, \& Tolias}]{Sinz2019}
Sinz, F.~H., Pitkow, X., Reimer, J., Bethge, M., \& Tolias, A.~S. (2019).
\newblock Engineering a less artificial intelligence.
\newblock {\em Neuron\/}, {\em 103\/}(6), 967--979.
\newline\urlprefix\url{https://doi.org/10.1016/j.neuron.2019.08.034}

\bibitem[{Sofroniew et~al.(2016)Sofroniew, Flickinger, King, \&
  Svoboda}]{Sofroniew2016-up}
Sofroniew, N.~J., Flickinger, D., King, J., \& Svoboda, K. (2016).
\newblock A large field of view two-photon mesoscope with subcellular
  resolution for in vivo imaging.
\newblock {\em Elife\/}, {\em 5\/}, e14472.

\bibitem[{Stringer et~al.(2019)Stringer, Pachitariu, Steinmetz, Reddy,
  Carandini, \& Harris}]{stringer2019spontaneous}
Stringer, C., Pachitariu, M., Steinmetz, N., Reddy, C.~B., Carandini, M., \&
  Harris, K.~D. (2019).
\newblock Spontaneous behaviors drive multidimensional, brainwide activity.
\newblock {\em Science\/}, {\em 364\/}(6437).

\bibitem[{Touryan et~al.(2005)Touryan, Felsen, \& Dan}]{Touryan2005}
Touryan, J., Felsen, G., \& Dan, Y. (2005).
\newblock Spatial structure of complex cell receptive fields measured with
  natural images.
\newblock {\em Neuron\/}, {\em 45\/}(5), 781--791.

\bibitem[{Ustyuzhaninov et~al.(2022)Ustyuzhaninov, Burg, Cadena, Fu, Muhammad,
  Ponder, Froudarakis, Ding, Bethge, Tolias, \& Ecker}]{Ustyuzhaninov2022}
Ustyuzhaninov, I., Burg, M.~F., Cadena, S.~A., Fu, J., Muhammad, T., Ponder,
  K., Froudarakis, E., Ding, Z., Bethge, M., Tolias, A.~S., \& Ecker, A.~S.
  (2022).
\newblock Digital twin reveals combinatorial code of non-linear computations in
  the mouse primary visual cortex.
\newline\urlprefix\url{https://doi.org/10.1101/2022.02.10.479884}

\bibitem[{Vintch et~al.(2015)Vintch, Movshon, \& Simoncelli}]{Vintch2015}
Vintch, B., Movshon, J.~A., \& Simoncelli, E.~P. (2015).
\newblock A convolutional subunit model for neuronal responses in macaque {V1}.
\newblock {\em J. Neurosci.\/}, {\em 35\/}(44), 14829--14841.

\bibitem[{Walker et~al.(2019)Walker, Sinz, Cobos, Muhammad, Froudarakis, Fahey,
  Ecker, Reimer, Pitkow, \& Tolias}]{Walker2019-oq}
Walker, E.~Y., Sinz, F.~H., Cobos, E., Muhammad, T., Froudarakis, E., Fahey,
  P.~G., Ecker, A.~S., Reimer, J., Pitkow, X., \& Tolias, A.~S. (2019).
\newblock Inception loops discover what excites neurons most using deep
  predictive models.
\newblock {\em Nat. Neurosci.\/}, {\em 22\/}(12), 2060--2065.

\bibitem[{Yamins et~al.(2014)Yamins, Hong, Cadieu, Solomon, Seibert, \&
  DiCarlo}]{Yamins2014}
Yamins, D. L.~K., Hong, H., Cadieu, C.~F., Solomon, E.~A., Seibert, D., \&
  DiCarlo, J.~J. (2014).
\newblock Performance-optimized hierarchical models predict neural responses in
  higher visual cortex.
\newblock {\em Proceedings of the National Academy of Sciences\/}, {\em
  111\/}(23), 8619--8624.
\newline\urlprefix\url{https://doi.org/10.1073/pnas.1403112111}

\bibitem[{Zhang et~al.(2018)Zhang, Lee, Li, Liu, Tang, Sing, Ming, Fang,
  Shiming, Lee, Li, Liu, \& Tang}]{Zhang2018-cs}
Zhang, Y., Lee, T.-S. T.~S., Li, M., Liu, F., Tang, S., Sing, T., Ming, L.,
  Fang, L., Shiming, L., Lee, T.-S. T.~S., Li, M., Liu, F., \& Tang, S. (2018).
\newblock Convolutional neural network models of {V1} responses to complex
  patterns.
\newblock {\em J. Comput. Neurosci.\/}, (pp. 1--22).

\bibitem[{Zipser \& Andersen(1988)}]{Zipser1988}
Zipser, D., \& Andersen, R.~A. (1988).
\newblock A back-propagation programmed network that simulates response
  properties of a subset of posterior parietal neurons.
\newblock {\em Nature\/}, {\em 331\/}(6158), 679--684.

\end{thebibliography}
\onecolumn
\newpage


\renewcommand{\figurename}{Supplemental Fig.}
\setcounter{figure}{0}

\section*{Supplementary Information} \label{note:FigS3b}

Supplemental Fig. 1 - Scan field location within primary visual cortex


\begin{figure*}[h!]
\centering
\includegraphics[width=0.95\linewidth]{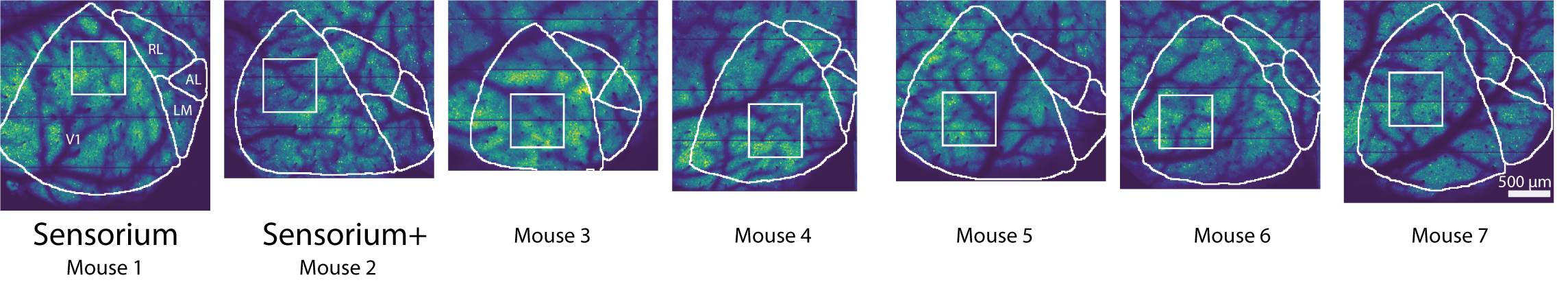}
\caption{\textbf{Scan field locations within primary visual cortex.}
Registered location of the scan field (white rectangle) to retinotopic mapping scan with manually annotated area boundaries (white lines) for primary visual cortex (V1) and anterolateral (AL), rostrolateral (RL), and lateromedial (LM) higher visual areas. The order of animals presented is identical to (Fig.~\ref{fig:Fig_2}c).
\label{fig:SFig_1}
}
\end{figure*}
\newpage

\end{document}